\documentclass[preprintnumbers,amsmath,amssymb,onecolumn]{revtex4}
\usepackage[driverfallback=dvipdfm]{hyperref}
\usepackage{color}
\usepackage{graphicx}
\usepackage{makeidx}
\usepackage{bm}
\usepackage[title]{appendix} 
\usepackage{tabu}
\usepackage{float}
\usepackage{chemarr}

\begin{document}
\title{Stochastic approach to study the properties of the complex patterns observed in cytokine and T-cells interaction process}
\author{Moirangthem Shubhakanta Singh$^1$, Mairembam Kelvin Singh$^1$ and R.K. Brojen Singh$^2$}
\email{brojen@jnu.ac.in}
\affiliation{$^1$Department of Physics, Manipur University, Canchipur-795003, Imphal, Manipur, India.\\ $^2$School of Computational $\&$ Integrative Sciences, Jawaharlal Nehru University, New Delhi-110067, India.}

\begin{abstract}
{\noindent}Patterns in complex systems store hidden information of the system which is needed to be explored. We present a simple model of cytokine and T-cells interaction and studied the model within stochastic framework by constructing Master equation of the system and solving it. The solved probability distribution function of the model show classical Poisson pattern in the large population limit $M,Z\rightarrow large$ indicating the system has the tendency to attract a large number small-scale random processes of the cytokine population towards the basin of attraction of the system by segregating from nonrandom processes. Further, in the large $\langle Z\rangle$ limit, the pattern transform to classical Normal pattern, where, uncorrelated small-scale fluctuations are wiped out to form a regular but memoryless spatiotemporal aggregated pattern. The estimated noise using Fano factor shows clearly that the cytokine dynamics is noise induced process driving the system far away from equilibrium.\\ 

\end{abstract}

%PACS number(s): 71.23, 72.15.R, 73.50
%%%%%%%%%%%%%%%%%%%%%%%%%%%%%%%%%%%%%%%%%%%%%%%%%%%%%%%%%%%%%%%%%%%%%%%%%%%%%%%%%%%%%%%%%%%%%%%%%%%%%%
\maketitle

\section{INTRODUCTION}
{\noindent}Patterns are inherent fundamental building blocks of complex systems. Each pattern is created by a self-organized certain population of interacting components of the system resulting aggregation having certain form of regularity, symmetry and structure which is the source of certain form of information and can participate in the system's collective decision making process \cite{Julia}. However, the whole process of each pattern may not be the exact sum of the individual activities of the constituting components which are nonlinear in characters, sensitive to the surrounding internal and external fluctuations and initial conditions, dissipative in nature, and the dynamics are generally strongly coupled \cite{Newman}. Each pattern is generally open in nature which allow exchange of energy, accessible states are far away from equilibrium, and the information flow is permitted across the boundary \cite{Werner}. The interaction among the components in a pattern could be of diverse form which drives the dynamics of individual components quite complicated in nature which could be the origin of complicated properties of each pattern \cite{Karaca}. Technically, the degree of pattern is generally characterized by complexity \cite{Cosma}. The complexity of the pattern can be measured by different techniques, some of which are information theoretical approach \cite{Hale}, statistical complexity measures \cite{Crutchfield}, Kolmogorov complexity \cite{Kolmogorov},  Lempel-Ziv complexity \cite{Aboy} etc. The study of the properties of these patterns may provide the information of how complex systems work at fundamental level.\\

{\noindent}The observation of the emergent properties of the patterns in the systems can be done mathematically using traditional central limit theorem by calculating large number of outcomes perceived from hidden processes of the distribution of the interacting random variables $\{x_i\};i=1,2,...,N$ corresponding to the aggregated $N$ components of the pattern \cite{Jaynes,Feller1,Feller2}. Depending on the observed properties of the probability distribution, the pattern can be in one of the classical patterns, namely, \textit{Norma, Poisson, Power law patterns etc} \cite{Frank} and these patterns, in most of the biological systems, are dynamic in nature depending on the fluctuations driving them and limiting distribution conditions \cite{Wootton}. If the spatiotemporal events are scattered randomly throughout the system, the probability distribution function becomes Poisson distribution and the pattern corresponds to classical \textit{Poisson pattern} \cite{Feller1,Feller2,Frank}. Since Poisson pattern is generally generated from the random processes smaller-scales, the basin of attraction of such patterns have aggregation of large scale small-scale random processes by segregating from nonrandom processes \cite{Kleiber}. On the other hand, classical \textit{Normal pattern} stores the information of mean $\langle x\rangle$ and variance $\sigma^2$ in the aggregation process of the components in the pattern $\mathcal{P}[\langle x\rangle,\sigma^2]$, where, sum of the uncorrelated small-scaled fluctuations of the components during the aggregation process cancels out to attract towards a certain form of spatiotemporal regular aggregated pattern \cite{Jaynes, Frank}. Hence, the dynamics of the aggregated components in such classical Normal patterns follow Brownian motion \cite{Uhlenbeck}. Further, if the aggregated pattern keeps only the information of mean, the pattern falls in \textit{exponential pattern} $\displaystyle\mathcal{P}[x]=\frac{1}{\langle x\rangle}e^{-\frac{x}{\langle x\rangle}}$, where, the pattern is contributed from the waiting time characterization of random events and the distribution has memoryless property \cite{Frank}. If the aggregation pattern preserves the information of geometric mean, the pattern follows classical \textit{Power law} pattern which can have fractal patterns which have self-similar process $\displaystyle\frac{\mathcal{P}[\alpha x]}{\mathcal{P}[x]}=\alpha^h=F(\alpha)$, where, $h$ is self-similarity exponent \cite{Mandelbrot,Barabasi}. From these probability distribution functions, one can measure hidden information stored in the respective patterns by calculating Shannon entropy, $\displaystyle 
\mathcal{H}=-\sum_{i}\mathcal{P}_i[x]ln\mathcal{P}_i[x]$ \cite{Shannon}. \\

{\noindent}The time evolution of the components in the patterns/systems is in general fluctuation driven stochastic process which can be well described in a probabilistic manner \cite{Kolmogorov}. The inherent fluctuations in the dynamics of the constituting variables arise from the internal (due to inter-molecular interaction) and external (due to fluctuations driven by surrounding the system i.e. temperature fluctuation, environmental fluctuations etc) fluctuations and have variety of roles in regulating the system \cite{Kampen,Swain}. The trajectory of the variables of the pattern/system and their realizations can be obtained using the Master equation, which is the discrete form of the Chapman-Kolmogorov equation \cite{Kolmogorov,Gardiner}, constructed from the information of the transitions of the states due to inter-variables interaction \cite{Kampen,Papoulis,McQuarrie}. On the other hand, computationally, the trajectory of each variables in the system can be traced using Monte carlo like stochastic simulation algorithm (SSA) due to Gillespie \cite{Gillespie}. The stochastic approach has wide variety of applications, ranging from microscopic biomolecules \cite{Bressloff} to macroscopic ecology \cite{Russell} to cosmology \cite{Kodama}, and then to finance \cite{Wolfgang,Michael}, cryptography \cite{Jonathan} etc. \\

{\noindent}Cytokines are certain category of diverse family of small proteins or glycoproteins that are secreted by specific cells of immune system and are produced throughout the body by cells of diverse embryological origin \cite{Jodie}. Cytokines provide the mechanism by which lymphocytes, inflammatory cells and haematopoietic cells can communicate with each other to mount and coordinate an effective immune response \cite{Dimitriou}. Interleukins (ILs) are the Cytokines secreted by leukocytes and affect the cellular responses of leucocytes \cite{Delgado}. Cytokines are pleiotropic which means that they have multiple effects and can increase both proliferation and death of a certain cell type \cite{Hart}. On the other hand, a pleiotropic cytokine that helps in the growth of T-cells is Interleukin-2 (IL-2) and intervenes in the activation-induced cell death (AICD) and causes differentiation of regulatory T-cells \cite{Wei}. Hence, interleukin-2 (IL-2) plays a major role in the proliferation of activated T cells \cite{Crawley}. These T-cells can identify and eliminate invading pathogens, and also attack human tissues \cite{Booki}. Further, a successful activation and differentiation of T-cells is aided by the presence of cytokines and specify the type of response that the T-cells give to a foreign antigen \cite{Anaya}. Now, it has been shown that, a particular type of T-cells known as the CD4$^+$ T-cell produces cytokine IL-2, which is responsible for both the proliferation and death of T-cells \cite{Refaeli,Li}. This cytokine can also be produced by T-cells, the T-cells proliferate to the cytokine again and are removed, they can also uptake and secrete the cytokine and the cytokine can then degrade itself \cite{Hart}. However, in the complicated regulating mechanism of cytokine and T-cells interaction what are dynamical properties of the system, how does the dynamics affect the patterns in the interacting system and how can we interrelate the emergent patterns to phenotyical behavior are still not answered fully. Further, what are the roles of the fluctuations in system's dynamics are still open questions.\\

{\noindent}In this work, we present a simple model of cytokine and T-cells interaction model. We then analyze the model using stochastic approach to understand how the probability distribution function of the cytokines aggregate to form different patterns at various situations. The results also explain the role of noise in regulating the system. Then some conclusions are made based on the results obtained.\\

\section{Methodology}
\subsection{Master Equation Formalism} 
{\noindent}The dynamics of stochastic particles can generally be described by The Chapman-Kolmogorov equation which is based on the probabilistic description of state transitions during the course of the particle's motion \cite{Kolmogorov,Gardiner}. This Chapman-Kolmogorov equation in integral form \cite{Gardiner} can relate the joint probability distributions of different sets of variables involved in a stochastic process where the transition of states of the system obey Markov process. If one consider a $'n'$-step transition probability of the stochastic process, then the jump probability $P(x_n,t_n|x_1,t_1)$ from the state $(x_1,t_1)$ to the state $(x_n,t_n)$ is given by,
\begin{eqnarray}
P(x_n,t_n|x_1,t_1)=\int ... \int dx_{n-1} ... dx_3dx_2P(x_n,t_n|x_{n-1},t_{n-1}|x_{n-2},t_{n-2})...P(x_2,t_2|x_1,t_1)
\end{eqnarray}
The standard way of deriving Chapman-Kolmogorov equation in differential form is to calculate the time evolution of the expectation value of a function $f(x)$ defined in the $n-$dimensional system given by,
\begin{eqnarray}
\frac{\partial}{\partial t}\langle f(x)\rangle=\int dxf(x)\frac{\partial}{\partial t}P(x,t|x^\prime,t^\prime)
\end{eqnarray}
Now, using the definition of differentiation to the partial derivative of the probability distribution function $P(x,t|x^\prime,t^\prime)$, expanding $f(x)$ with Taylor series expansion and then rearranging the terms by keeping terms upto second order derivative, one can get the following Chapman-Kolmogorov equation \cite{Kolmogorov,Gardiner},
\begin{eqnarray}
\label{cke}
\frac{\partial}{\partial t}P(x,t|x^\prime,t^\prime)&=&-\sum_i \frac{\partial}{\partial u_i}\left[A_i(u,t)P(u,t|x^\prime,t^\prime)\right]+\frac{1}{2}\sum_{i,j}\frac{\partial^2}{\partial u_i\partial u_j}\left[B_{ij}(u,t)P(u,t|x^\prime,t^\prime)\right]\nonumber\\
&&+\int dx \left[W(u|x,t)P(x,t|x^\prime,t^\prime)-W(x|u,t)P(u,t|x^\prime,t^\prime)\right]
\end{eqnarray}
where, $\left\lbrace W\right\rbrace$ are Wiener functions which are transition probabilities of the states and $A_i$ and $B_{ij}$ are the functions which depend the properties of the system. The continuum approximation of the equation \eqref{cke}, which is contributed from the first two terms in the right hand side, provides Fokker-Planck-Kolmogorov equation \cite{Gardiner}. The Master equation is generally the discrete form of the Chapman-Kolmogorov equation \cite{Kolmogorov,Gardiner, Kampen} which is contributed from the jump process of the particle dynamics (keeping the last two terms in the right hand side), which is given by,
\begin{eqnarray}
\label{me}
\frac{\partial}{\partial t}P(x,t|x^\prime,t^\prime)=\int dx \left[W(u|x,t)P(x,t|x^\prime,t^\prime)-W(x|u,t)P(u,t|x^\prime,t^\prime)\right]
\end{eqnarray}
This equation can also be expressed by replacing integration by summation over configurational states indicating jump process \cite{Gardiner,Kampen,Bressloff} which is easier to handle and more directly related to physical systems \cite{Kampen}. Master equation can also be seen as the time evolution of the configurational probability for Markov processes of systems that jump from one to another state which is due to birth and death processes of the particles in the system. There are different techniques to solve Master equation, namely, using generating function technique \cite{McQuarrie}, Fourier transform technique \cite{Kodama}, Laplace transform technique \cite{McNeil}, Laplace-Fourier transform technique \cite{Scalas} etc. However, solving Master equation of complex systems involving multi-variable is quite difficult except for simple systems.

\subsection{Generating function technique to solve Master equation}
{\noindent}The generating function (GF) technique is generally used to solve Master equations for simple low dimensional systems \cite{McQuarrie}. Brief procedure to solve Master equation using GF is given as follows. The $n-$dimensional GF $G(s_1,s_2,...,s_n)$ is the transformation map of the probability distribution function $P(x_1,x_2,...,x_n;t)$ as defined by,
\begin{eqnarray}
\label{gft}
G(s_1,s_2,...,s_n)=\sum_{x_1,x_2,...,x_n}\prod_{i=1}^n s_i^{x_i}P(x_1,x_2,...,x_n;t)
\end{eqnarray}
Now, multiplying the above Master equation \eqref{me} by $\sum_{x_1,x_2,...,x_n}\prod_{i=1}^n s_i^{x_i}$ and substituting equation \eqref{gft}, one can transform the Master equation to spatiotemporal partial differential equation in the GF. The partial differential equation in $G$ can be solved using the boundary condition at $t=0$, 
\begin{eqnarray}
G(s_1,s_2,...,s_n;0)=\prod_{i=1}^n s_i^{N_i^{[0]}}
\end{eqnarray}
where, $\{N_i^{[0]}\};i=1,2,...,n$ is the set of initial populations of the respective variables $x_1,x_2,...x_n$. Now, the solution $G$ is put back to the GF \eqref{gft} and equating the coefficients of the terms containing $\prod_{i=1}^n s_i^{x_i}$ from both sides, one can arrive at the functional form of $P(x_1,x_2,...,x_n)$ which is the solution of Master equation \eqref{me}. However, solving Master equation for any high dimensional system is quite difficult and is still an open question.\\

{\noindent}The generating function also allows us to determine the observables which will describe the properties of the system, such as, the role of fluctuations in the system and many others. The analysis of GF and probability distribution function could reveal various other hidden patterns and their properties of the system.\\

{\noindent}The observables of the system namely, mean, variance and other measurable parameters can be determined using the generating function. The mean of any variable $x_i$ can be obtained from the following relation,
\begin{eqnarray}
\langle x_i \rangle&=&\left.\frac{\partial G(s_i,t)}{\partial s_i}\right|_{s_i=1;i=1,2,...,n}\\
\langle x_i^2 \rangle&=&\left.\frac{\partial^2 G(s_i,t)}{\partial s_i^2}\right|_{s_i=1;i=1,2,...,n}+\langle x_i \rangle\\
\sigma_i^2&=&\langle x_i^2 \rangle-\langle x_i \rangle^2
\end{eqnarray}
The fluctuations involved in the dynamics of the particles in the system can be well quantified by measuring \textit{Fano factor} $F$ of the variable corresponding to the particle which is given by \cite{Fano}, 
\begin{eqnarray}
F=\frac{\sigma_i^2}{\langle x_i \rangle}
\end{eqnarray}
The Fano Factor is an important parameter which can determine the role of fluctuations or noise in the dynamics of the particles of the system \cite{Kamil}. Depending on the values of $F$, one can determine the behavior of the particle which undergo various processes \cite{Kamil,Chanu}.
\begin{itemize}
\item If $F=1$, then, the process is a Poisson process and the particles in the system undergo Brownian motion. In such situation, there is no specific role of the fluctuations in the dynamics of the system.
\item If $F>1$, the process is a noise-enhanced process and is also known as super-Poisson process. Here, the fluctuations has significant role in regulating the dynamics of the system.
\item If $F<1$, the process is called a sub-Poisson process. Here, the noise has no significance but still has a tendency to regulate the dynamics of the system.
\end{itemize}

\subsection{Cytokines and T-cell interaction model}
{\noindent}The proliferation of T-cells (T) involves the production of a cytokine $IL-2$ (Z), and this cytokine plays a significant role in the T-cell proliferation and death \cite{Hart}. In this simple model, if we denote the rate of proliferation as $X$ and rate of death as $Y$, and when these two rates are functions of the produced cytokine $Z$, we can write the concentration of the cytokine $Z$ at any instant of time by a simple first order differential equation \cite{Hart},
\begin{eqnarray}
\label{xy}
\frac{dZ}{dt}=X(Z)T-Y(Z)Z
\end{eqnarray}
The maximum value of $Z$ during the process can be obtained by calculating the fixed point of the equation \eqref{xy}, $\frac{dZ}{dt}=0$ giving the value, $Z^*=\frac{X^*(Z^*)}{Y^*(Z^*)}T$. This steady state population of the cytokine $Z$ is generally independent of its initial population at time $t=0$. However, the initial population of cytokine trigger the T-cell proliferation by producing another type of cytokine in response \cite{Hart}. Moreover, the cytokine IL-2 can be produced from other external sources other than T-cell proliferation. Further, T-cells can uptake and secrete the cytokine IL-2 which are needed to be incorporated to the model.\\

{\noindent}Considering various processes involved in the proliferation of T-cells and its death with the production of cytokine IL-2, the above model \eqref{xy} can be represented by following form of the equation \cite{Hart},
\begin{eqnarray}
\label{model}
\frac{dZ}{dt}=k_1+k_2T-k_3TZ-k_4
\end{eqnarray}
where $k_1$ is the secretion rate by an external source, $k_2$ is the secretion rate by T-cells, $k_3$ is the rate of uptake of Z by T-cells and $k_4$ is the rate of degradation of Z by itself. This equation can translated into the following four reactions,
\begin{eqnarray}
\label{re}
\alpha\stackrel{\rm k_1}{\longrightarrow} Z;~~
T\stackrel{\rm k_2}{\longrightarrow} Z;~~
T+Z\stackrel{\rm k_3}{\longrightarrow} T;~~
Z\stackrel{\rm k_4}{\longrightarrow} \phi
\end{eqnarray}
\begin{figure}
\label{Fig. (i)}
\begin{center}
\includegraphics[height=5.0cm,width=8.0cm]{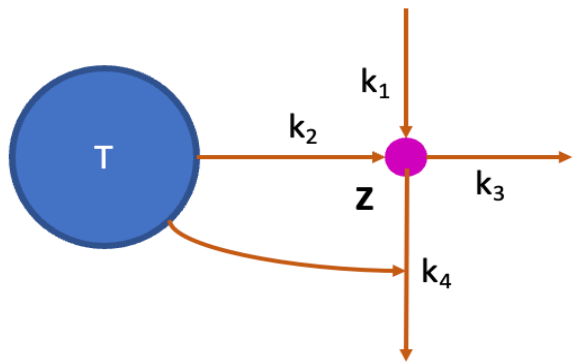}
\caption{Schematic diagram representing T-cell and cytokine IL-2 (Z) interaction mechanism.}
\end{center}
\end{figure}
{\noindent}The first reaction represents the secretion of cytokine $Z$ by an external source $\alpha$, the second reaction shows the secretion of cytokine $Z$ by T-cells, the third reaction indicates uptake of cytokine $Z$ by T-cells and the fourth reaction represents the degradation of cytokine $Z$. The schematic diagram of these reactions is given in Figure 1.

\section{Results}

\subsection{Stochastic approach to study cytokine and T-cells interaction model}
{\noindent}The study of the stochastic dynamics of the system can lead us to explore the role of fluctuations and other factors which regulate the behavior of the system. Since the patterns exhibited in the interaction of the cytokine and T-cells at any instant of time is random and probabilistic in nature, understanding the properties of the patterns within the stochastic framework could be important to correlate with the exhibited phenotypic behavior. We consider the equivalent reaction set \eqref{re} of the model equation \eqref{model} and assume that the total population of cytokine $Z$ and T-cells is a constant, i.e, $T+Z=constant=M$. Further, as the T-cells recruits cytokine in a quite fast rate \cite{Rogers} and the T cells population in our immune system always tends to an equilibrium/steady state \cite{Almeida}, the population of T-cells is also taken to be a constant as compared to cytokine population in our calculation.\\

{\noindent}Consider the transition of states given by the model reaction set \eqref{model} obey the Markov property \cite{Jodie}, which means that the future state of a system depends only on the present state and not on any of the past states, we can determine the transition probabilities from one state to the next for each reaction using the Law of mass action. From the model \eqref{model}, the reaction $\alpha\stackrel{\rm k_1}{\longrightarrow} Z$ indicates the production of cytokine Z by T-cells, where, $\alpha$ is a constant source. Since the firing of the reaction follow Markov process, the transition rates of the process can be calculated using mass action law in the state change processes and are given by, $\omega_1=\alpha k_1\Delta t+O(\Delta t^2)$ and $\omega_1^\prime=\alpha k_1\Delta t+O(\Delta t^2)$ respectively. Next, the reaction $T\stackrel{\rm k_2}{\longrightarrow} Z$ reveals the proliferation of T-cells to recruit cytokine Z and removal of itself with a rate $k_2$. The corresponding transition rates are given by,
$\omega_2=k_2\beta\Delta t+O(\Delta t^2)$ and $\omega_2^\prime=k_2\beta\Delta t+O(\Delta t^2)$ respectively, where, $\beta$ is the mean value of population of T-cells as its fluctuation is approximated to be negligible.
Further, the reaction $T+Z\stackrel{\rm k_3}{\longrightarrow} T$ shows that the T-cells can also uptake a cytokine Z, secrete another cytokine Z and then remove itself. This means that the state of cytokine Z remains constant at state Z. The transition rates of this reaction are given by, $\omega_3=k_3\beta(Z+1)\Delta t+O(\Delta t^2$ and $\omega_3^{\prime}=k_3\beta Z\Delta t+O(\Delta t^2$ respectively. Lastly, the degradation of cytokine Z is indicated by the reaction $Z\stackrel{\rm k_4}{\longrightarrow} \phi $. The transition rates are given by, $\omega_4=k_4(Z+1)\Delta t+O(\Delta t^2)$ and $\omega_4^{\prime}=k_4Z\Delta t+O(\Delta t^2)$ respectively.\\

{\noindent}Now, according to the principle of detailed balance, the process of transition from one state to another should be in equilibrium with the corresponding reverse process of transition. Using this principle, we can determine the probability, $P(T, Z; t+\Delta t)$, that the system is in the state $(T,Z)$ at any instant of time $t+\Delta t$ provided that the system was in some other state at time $t$. Now, using the transition rates for each reaction, we can arrive at the following detailed balance condition,
\begin{eqnarray}
P(Z;t+\Delta t)&=&\omega_1 P(Z-1,t)+\omega_2 P(Z-1,t)+\omega_3 P(Z+1,t)\nonumber\\
&&+\omega_4P(Z+1,t)+[1-(\omega_1^\prime+\omega_2^\prime+\omega_3^\prime+\omega_4^\prime)]P(Z,t)\nonumber\\
&=& \alpha k_1P(Z-1; t)\Delta t+\beta k_2P(Z-1, t)\Delta t\nonumber\\
&& +\beta k_3(Z+1) P(Z+1, t)\Delta t+k_4 (Z+1)P(Z+1, t)\Delta t\nonumber\\
&& +[1-(\alpha k_1+\beta k_2+\beta k_3Z+k_4Z)\Delta t]P(Z,t)
\end{eqnarray}
Now, rearranging the above equation and using the definition of differentiation in the limit $\Delta t \rightarrow 0$, we arrived at the following Master equation,
\begin{eqnarray}
\label{master}
\frac{\partial P(Z;t)}{\partial t}&=&\alpha k_1 P(Z-1,t)+\beta k_2 P(Z-1,t)+\beta k_3(Z+1)P(Z+1,t)\nonumber\\
&&+k_4(Z+1) P(Z+1, t)-[\alpha k_1+\beta k_2+\beta k_3Z+k_4Z]P(Z,t)
\end{eqnarray}
This Master equation \eqref{master} can be solved using generating function technique which we have explained in the methodology section. The generating function in one dimension can be defined as.
\begin{eqnarray}
G(s,t)=\sum_Z s^Z P(Z,t)
\end{eqnarray}
with the initial condition, $G(s,0)=S^N$ at $t=0$, where $N$ is the original number of cytokine population in time $t=0$. Now, multiply equation \eqref{master} by $s^Z$ on both sides, summing over Z, and after rearranging the terms, we have,
\begin{eqnarray}
\label{gpde}
\frac{\partial G(s,t)}{\partial t}=-(\beta k_3+k_4)(s-1)\frac{\partial G(s,t)}{\partial s}+(\alpha k_1+\beta k_2)(s-1)G(s,t)
\end{eqnarray}
This spatiotemporal partial differential equation \eqref{gpde} can be solved using standard Lagrange's Charpit characteristic method for partial differential equations \cite{Delgado}. Now, applying this method, the solution can be obtained from the following set of equations,
\begin{eqnarray}
\label{lce}
\frac{dt}{1}=\frac{ds}{(\beta k_3+k_4)(s-1)}=\frac{\partial G(S,t)}{(\alpha k_1+\beta k_2)(s-1)G(s,t)}
\end{eqnarray}
Now, let us consider the first equation constructed from the above set of equations \eqref{lce},
$dt=\frac{ds}{(\beta k_3+k_4)(s-1)}$. Integrating both sides and rearranging the terms, we get,
\begin{eqnarray}
\label{l1}
\lambda_1=e^{C_1}=(s-1)e^{-(\beta k_3+k_4)t}
\end{eqnarray}
where, $\lambda_1$ is a constant and $C_1$ is the constant of integration.
Next, we consider second equation from the set of the equations eqref{lce}, given by, $\frac{\partial G(S,t)}{(\alpha k_1+\beta k_2)(s-1)G(s,t)}=\frac{ds}{(\beta k_3+k_4)(s-1)}$. Now integrating and rearranging the terms, we have,
\begin{eqnarray}
\label{l2}
G(s,t)=\lambda_2 e^{\left[\frac{\alpha k_1+\beta k_2}{\beta k_3+k_4}\right]s}
\end{eqnarray}
where, $\lambda_2=e^{C_2}$ is a constant and $C_2$ is another constant of integration. Now, combining the equations \eqref{l1} and \eqref{l2}, we can write $G$ in the following form,
\begin{eqnarray}
\label{g}
G(s,t)=F\left[(s-1)e^{-(\beta k_3+k_4)t}\right]e^{\left[\frac{\alpha k_1+\beta k_2}{\beta k_3+k_4}\right]s}
\end{eqnarray}
Now, to find out the functional form of the function $F$ in equation \eqref{g}, we take the initial boundary condition given by,
\begin{eqnarray}
G(s,0)=s^M
\end{eqnarray}
where, $M$is the initial cytokine population at $t=0$. Putting this condition to equation \eqref{g}, we have the $F$ given by,
\begin{eqnarray}
F(s)=(s+1)^Me^{\left[-\frac{\alpha k_1+\beta k_2}{\beta k_3+k_4}\right]s}
\end{eqnarray}
Then, using this equation to the equation \eqref{g}, we have the GF given by,
\begin{eqnarray}
\label{gf}
G(s,t)=e^{ab}\left[(s-1)a+1\right]^M e^{bs(1-a)}
\end{eqnarray}
where, $a=e^{-(\beta k_3+k_4)t}$ and $b=\frac{\alpha k_1+\beta k_2}{\beta k_3+k_4}$.
Now, rearranging the first factor of the equation \eqref{gf} in the form $F_1=\left[(s-1)e^{-(\beta k_3+k_4)t}+1\right]^M=a^M\left[s+\frac{1-a}{a}\right]^M$, and using Binomial expansion, we have, $F_1=\displaystyle\sum_{i=0}^{M}~^MC_i s^i\left(\frac{1-a}{a}\right)^{M-i}$. The second factor in the equation \eqref{gf} can be rearranged and expand in the form, $e^{bs(1-a)}=\displaystyle\sum_{j=0}^{\infty}\frac{[b(1-a)]^j}{j!}s^j$. Now putting the factors back to the equation \eqref{gf}, and taking $Z=i+j$, we have,
\begin{eqnarray}
\label{gff}
G(s,t)&=&e^{ab}\sum_{i=0}^{M}\sum_{j=0}^{\infty}~^MC_i\frac{\left[b(1-a)\right]^j}{j!}s^{i+j}\left[\frac{1-a}{a}\right]^{M-i}a^M\nonumber\\
&=&e^{ab}\sum_{Z}s^{Z}\sum_{i=0}^{M}~^MC_i\frac{b^{Z-i}}{(Z-i)!}\left(1-a\right)^{M-2i+Z}a^i\nonumber\\
&=&\sum_{Z}s^ZP(Z,t)
\end{eqnarray}
Equating the coefficient of $s^Z$, we arrived at the expression for $P(Z,t)$, given by,
\begin{eqnarray}
\label{pd}
P(Z,t)=e^{ab}\sum_{i=0}^{M}~^MC_i\frac{b^{Z-i}}{(Z-i)!}\left(1-a\right)^{M-2i+Z}a^{i}
\end{eqnarray}
Now, let us study the role of noise in cytokine dynamics triggered by T-cells which can be analyzed by calculating Fano factor \cite{Fano}. To calculate the Fano factor $\mathcal{F}$ of the cytokine dynamics we need to calculate observables, mean $\langle Z\rangle$ and variance $\sigma^2=\langle Z^2\rangle-\langle Z\rangle^2$. Using equation eqref{gf}, the mean of the cytokine dynamics is given by,
\begin{eqnarray}
\label{mean}
\langle Z \rangle&=&\left.\frac{\partial G(s,t)}{\partial s}\right|_{s=1}\nonumber\\
&=&e^{b}\left[a(M-b)+b\right]
\end{eqnarray}
The asymptotic behaviors of the mean can be calculated as follows. Taking $\displaystyle\lim_{t\rightarrow 0}\langle Z \rangle=e^{b}M$, which indicates that the initial population of cytokine $Z$ depends on various interaction processes involved in the model i.e. various rate constants and population of T-cells $\beta$. Since the total population has to be $M$, we should have $b=0$, which means that $\frac{\alpha k_1+\beta k_2}{\beta k_3+k_4}=0$ i.e. $\alpha k_1+\beta k_2=0$ revealing that the overall production of cytokine both from external sources and secretion by T-cell should be zero.
On the other hand, since, $\displaystyle\lim_{t\rightarrow\infty}\langle Z \rangle\rightarrow be^{b}=constant$ indicating the population of cytokine reach a stable constant value at significantly large time.
Next, the variance is given by,
\begin{eqnarray}
\label{sig}
\sigma^2&=&\langle Z^2 \rangle-\langle Z\rangle^2=\left.\frac{\partial^2G(s,t)}{\partial s^2}\right|_{s=1}+\langle Z\rangle-\langle Z\rangle^2\nonumber\\
&=&e^b\left[(Ma-ab+b)^2-Ma^2\right]+\langle Z\rangle-\langle Z\rangle^2
\end{eqnarray}
Now, the Fano factor is calculated as follows,
\begin{eqnarray}
\label{fano}
\mathcal{F}&=&\frac{\sigma^2}{\langle Z \rangle}\nonumber\\
&=&1+\langle Z \rangle\left(e^{-b}-1\right)-e^b\frac{Ma^2}{\langle Z \rangle}
\end{eqnarray}
Now, we analyze the Fano factor to understand the behavior of the cytokine dynamics.\\

\noindent\textbf{Proposition 1}
\noindent\textit{The asymptotic behavior of the cytokine dynamics allows only sub-Poissonian process.}\\

\noindent\textbf{Proof:} \textit{Let us calculate the asymptotic values of Fano factor $\mathcal{F}$ from equation \eqref{fano}.}
\begin{eqnarray}
\lim_{t\rightarrow 0}\mathcal{F}=M\left(1-e^b\right)
\end{eqnarray}
\textit{From the expression of $b$, $b\ne 0$ and $b\rangle 0$, such that, $e^b\rangle 1$. Hence, $\displaystyle\lim_{t\rightarrow 0}F\rightarrow -ve$ which can not be physically reliable and can not say anything about the cytokine dynamics. On the other hand,}
\begin{eqnarray}
\lim_{t\rightarrow\infty}\mathcal{F}=1-b\left(e^b-1\right)< 1
\end{eqnarray}
\textit{This indicates that the cytokine process follows sub-Poissonian nature. The involved noise in the cytokine dynamics tries to stabilize its dynamical behavior.}\\

\noindent\textbf{Proposition 2}
\noindent\textit{At any instant of time $t~0\langle t\langle\infty$, the cytokine dynamics noise-enhanced process.}\\

\noindent\textbf{Proof:} \textit{The cytokine dynamics becomes noise enhanced process or super-Poissonian process when $\mathcal{F}$ satisfies $\mathcal{F}>1$. Applying this condition in equation \eqref{fano}, the condition becomes,}
\begin{eqnarray}
\label{ff}
\langle Z\rangle^2>\frac{e^{2b}Ma^2}{1-e^b}
\end{eqnarray}
\textit{Since, $1\le e^b<\infty$, $1-e^b<0$. This indicates that left hand side term of equation \eqref{ff} is negative value. Since, $\langle Z\rangle^2>0$, the condition \eqref{ff} is always satisfied. Further, since $\langle Z\rangle^2$ can not be negative the process can not be Poissonian and sub-Poissonian process. Hence, only possible role of noise for $0\langle t\langle\infty$ is noise enhanced process.}\\

\subsection{Patterns observed in cytokines distribution dynamics}
{\noindent}The simple cytokine and T-cells interaction model exhibits complicated system's dynamics where the role of noise is quite significant in regulating the dynamics. Interesting question is even though the dynamics of the cytokine molecules are complicated they should follow some traditional equation of motion. Since the typical size of the single cytokine molecule ($IL2$) is around $450nm$ in diameter \cite{Bakker} which is within mesoscopic scale size \cite{Salje}, they are very sensitive and significantly affected by internal and external fluctuations \cite{Ritort}. We intend to study the patterns of the distribution functions these particles by analyzing the probability distribution function given by equation \eqref{pd}.\\

\noindent\textbf{Theorem 1}
\noindent\textit{Near thermodynamic limit, $M,Z\rightarrow\infty$ and $\langle Z\rangle\rightarrow finite$, }
\begin{itemize}
\item \textit{The cytokine distribution function $P(Z,t)$ follows modified Poisson distribution:}
\begin{eqnarray}
\label{pois}
P(Z,t)\sim e^{-a(M-b)+be^{-a\left[1-\frac{M}{b}\right]}}\times Pois\left[be^{-a\left(1-\frac{M}{b}\right)}\right]
\end{eqnarray}
\textit{which is independent of cytokine population variable $Z$.} 
\item \textit{The information stored in classical Poisson pattern is given by,}
\begin{eqnarray}
\mathcal{I}\sim e^{-a(M-b)+\Gamma}\times ln\left[\frac{M}{\sqrt{2\pi e\Gamma}}\right]
\end{eqnarray}
\end{itemize}

\noindent\textbf{Proof:} \textit{At large $M$ limit, the equation \eqref{pd} can be written as,}
\begin{eqnarray}
\label{pois1}
P(Z,t)&=&e^{ab}\sum_{i=0}^{M}\frac{\left[b(1-a)\right]^{Z-i}}{(Z-i)!}\times\left[~^MC_i a^i(1-a)^{M-i}\right]\nonumber\\
&\sim&e^{ab}\sum_{i=0}^{M}\frac{\left[b(1-a)\right]^{Z-i}}{(Z-i)!}\times\left[\frac{\mu^i e^{-\mu}}{i!}\right];~~~\mu=Ma;~~M\rightarrow large\nonumber\\
&\sim&e^{ab}\sum_{i=0}^{M}\frac{M^ie^{-\mu}b^{Z-i}}{Z!}\times~^ZC_ia^i(1-a)^{Z-i}\nonumber\\
&\sim&e^{ab}\sum_{i=0}^{M}\frac{M^ie^{-\mu}b^{Z-i}}{Z!}\times\frac{\epsilon^ie^{-\epsilon}}{i!}~~~\epsilon=Za;~~Z\rightarrow large\nonumber\\
&\sim&e^{ab}\frac{e^{-(\mu+\epsilon)}b^Z}{Z!}\sum_{i=0}^{M}\frac{\left[\frac{MZa}{b}\right]^i}{i!}\nonumber\\
&\sim&e^{ab}\frac{e^{-(\mu+\epsilon)}b^Z}{Z!}\times e^{\frac{MZa}{b}}\nonumber\\
&\sim&e^{-a(M-b)+be^{-a\left(1-\frac{M}{b}\right)}}\times Pois\left[be^{a\left(\frac{M}{b}-1\right)}\right]
\end{eqnarray}
\textit{The Poisson distribution is independent of $Z$ indicating the cytokines' distribution becomes homogeneous approximately everywhere in the system when the population of $Z$ is significantly large.}\\

\noindent\textit{Now, the complexity in the distribution of the cytokine molecules whose dynamics undergo stochastic process can be calculated using Shannon entropy $\mathcal{H}$ using the modified Poisson probability distribution function $\mathcal{P}(\Gamma,t)=Pois(\Gamma,t)$ given by equation \eqref{pois1}, where, $\Gamma=be^{a\left(\frac{M}{b}-1\right)}$. We follow the procedure of calculation given in the work \cite{Evans}. Then the terms are rearranged to get the expression for $\mathcal{H}$ as in the following,}
\begin{eqnarray}
\label{h1}
\mathcal{H}&=&-\sum_{u=0}^{M}\mathcal{P}(\Gamma\rightarrow u)ln[\mathcal{P}(\Gamma\rightarrow u)]\nonumber\\
&=&e^{-a(M-b)+\Gamma}\times\left[\sum_{u=0}^{\infty}\mathcal{P}(u)ln\left(\frac{\Gamma^ue^{-\Gamma}}{u!}\right)  \right]\nonumber\\
&=&e^{-a(M-b)+\Gamma}\times\left[\Gamma-\Gamma ln\Gamma+\sum_{u=0}^{\infty}\mathcal{P}(u)ln\left(u!\right)  \right]
\end{eqnarray}
\textit{Then substituting the series expression for $ln(i!)$ given by the Malmsten's representation \cite{Erdilyi} as in the following,}
\begin{eqnarray}
\label{u}
ln(u!)=\int_{0}^{\infty}dx\left[u-\frac{1-e^{-ux}}{1-e^{-u}}\right]\frac{e^{-x}}{x}
\end{eqnarray}
\textit{The equation \eqref{h1} becomes,}
\begin{eqnarray}
\label{h2}
\mathcal{H}&=&e^{-a(M-b)+\Gamma}\times\left[\Gamma-\Gamma ln\Gamma+\int_{0}^{\infty}dx\frac{e^{-x}}{x}\left(\Gamma-\frac{1-\langle e^{-ux}\rangle}{1-e^{-x}} \right)   \right]\nonumber\\
&=&e^{-a(M-b)+\Gamma}\times\left[\Gamma-\Gamma ln\Gamma+\int_{0}^{\infty}dx\frac{e^{-x}}{x}\left(\Gamma-\frac{1-e^{-x(1-e^{-x})}}{1-e^{-x}} \right)   \right]
\end{eqnarray}
\textit{The integration in equation \eqref{h2} can be done using series of formulae listed \cite{Erdilyi}, doing integration by parts and rearranging the terms, we get the following expression for entropy \cite{Evans},}
\begin{eqnarray}
\label{hm}
\mathcal{H}&\sim&e^{-a(M-b)+\Gamma}\times\left[\frac{1}{2}ln(2\pi e\Gamma)-\sum_{i=0}^{\infty}c_i(i-1)!\Gamma^{-i}\right]\nonumber\\
&\sim&\frac{1}{2}e^{-a(M-b)+\Gamma}\times ln(2\pi e\Gamma)~~(\Gamma\rightarrow\infty)
\end{eqnarray}
\textit{Now, the information stored in the Poisson pattern of the cytokine population can be calculated using the following expression for information $\mathcal{I}$ due to the surprises given by the Shannon entropy given by equation \eqref{h1},}
\begin{eqnarray}
\label{I}
\mathcal{I}=\mathcal{H}_{max}-\mathcal{H}
\end{eqnarray}
\textit{where, $\mathcal{H}_{max}$ is the maximum entropy which can be calculated using maximum entropy principle \cite{Guiasu}. The main procedure of this principle is to define a function $\Omega=\mathcal{H}-\kappa\Phi$, where, $\kappa$ is the Lagrange's multiplier, and $\Phi$ is a function constructed from the constraint $\displaystyle\sum_{i=0}^{M}\mathcal{P}_i=1$, such that, $\Phi=\displaystyle\sum_{i=0}^{M}\mathcal{P}_i-1$. The Shannon entropy given in equation \eqref{h1} can also be written as $\mathcal{H}=-\displaystyle\sum_{i=0}^{M}\mathcal{P}_iln\mathcal{P}_i$. Now the function $\Omega$ becomes,}
\begin{eqnarray}
\Omega=-\sum_{i=0}^{M}\mathcal{P}_iln\mathcal{P}_i-\kappa\left[\sum_{i=0}^{M}\mathcal{P}_i-1\right]
\end{eqnarray}
\textit{Taking the derivative of the function $\Omega$ with respect to $\mathcal{P}_i$, and putting $\displaystyle\frac{\partial\Omega}{\partial\mathcal{P}_i}=0$, and solving for $\mathcal{P}_i$ from the resulting equation, we have,
$\displaystyle\mathcal{P}_i=e^{-[1+\kappa]}$
The Lagrange's multiplier $\kappa$ can be calculated by putting this expression to the condition, $\displaystyle\sum_{i=0}^{M}\mathcal{P}_i=1$, and it is found that $\displaystyle\mathcal{P}_i=\frac{1}{M}$ for all values of $i$. Now, putting this expression to equation \eqref{h1}, we get, $\displaystyle\mathcal{H}_{max}=e^{-a(M-b)+\Gamma}\times ln[M]$. Now, putting this expression to the equation \eqref{I}, and using the equation \eqref{hm}, we get,}
\begin{eqnarray}
\label{ih}
\mathcal{I}\sim e^{-a(M-b)+\Gamma}\times ln\left[\frac{M}{\sqrt{2\pi e\Gamma}}\right]
\end{eqnarray}
\textit{This equation \eqref{ih} indicates that $\mathcal{I}=\mathcal{I}(t)$. From the asymptotic values of $\mathcal{I}$, we get that, (i) at large time limit, ($t\rightarrow\infty$), $\displaystyle\mathcal{I}_\infty=\lim_{t\rightarrow\infty}\mathcal{I}\sim e^{b}
\times ln\left[\frac{M}{\sqrt{2\pi eb}}\right]$, indicating if the overall production rate of the cytokine ($\alpha k_1+\beta k_2$) is faster than overall decay ($\beta k_3+k_4$) then, $b\gg 1$, and hence, $\mathcal{I}_0$ will become small. This condition indicate that the probability of finding of cytokines will be increased due to the increase in the population of cytokines, and hence, information stored will be decreased as the system goes to more ordered state. However, if overall production rate of the cytokines is smaller than the overall decay rate then $b\ll 1$ indicating $\mathcal{I}_0$ increases significantly indicating the system goes more disorder state. On the other hand, (ii) at initial time $t=0$, $\mathcal{I}_{0}=\displaystyle\lim_{t\rightarrow 0}\mathcal{I}=e^{-(M-b)+be^{\frac{M}{b}-1}}\times ln\left[\frac{M}{\sqrt{2\pi be^{\frac{M}{b}}}}\right]$.
}\\

\noindent\textbf{Theorem 2}
\noindent\textit{Near thermodynamic limit, and at the large mean value $\Gamma\rightarrow\infty$ (overall mean cytokine production rate is much higher than overall mean decay rate: $b\gg 1$),}
\begin{itemize}
 \item\textit{ The distribution approaches to classical Normal pattern:}
\begin{eqnarray}
\label{normal}
P(Z,t)\sim e^{-a(M-b)+\Gamma}\times \mathcal{N}\left[\Gamma,\Gamma\right]
\end{eqnarray}
\item\textit{The information stored in Normal pattern is given by,}
\begin{eqnarray}
\mathcal{I}=e^{-a(M-b)+\Gamma}\left[ln(M)+a(M+b)-\Gamma-\frac{d_1ln(\Gamma\sqrt{2\pi})+d_2}{\Gamma\sqrt{2\pi}} \right]
\end{eqnarray}
\end{itemize}

\noindent\textbf{Proof:} \textit{The proof is straight forward. Near thermodynamic limit, $M,Z\rightarrow large$ $P(Z,t)$ becomes Poisson distribution given by \eqref{pois1}. Further, taking $\Gamma\rightarrow large$, where, $\Gamma=be^{a\left(\frac{M}{b}-1\right)}$ and using Stirling's formula, $Z!\approx Z^Ze^{-Z}\sqrt{2\pi Z}$ to the Poisson distribution, $\displaystyle Pois(\Gamma)=\frac{\Gamma^Ze^{-\Gamma}}{Z!}$, we expand it around the mean $\Gamma$ and get Normal distribution, $\displaystyle Pois(\Gamma)\approx\mathcal{N}(\Gamma,\Gamma)$. This proves equation \eqref{normal}. 
}\\

\noindent\textit{Now, let us calculate the information stored in the classical normal pattern which can be calculated using the equation \eqref{I}. The Shannon entropy, which can be used to calculate the complexity in complex system, can be obtained by using the equation \eqref{normal} and rearranging the terms as given below, }
\begin{eqnarray}
\label{nh}
\mathcal{H}&=&-\sum_{u=0}^{M}\mathcal{P}(\Gamma\rightarrow u)ln\mathcal{P}(\Gamma\rightarrow u)\nonumber\\
&=&-e^{-a(M-b)+\Gamma}\times\left[\sum_{u=0}^{M}\mathcal{N}(\Gamma\rightarrow u)ln\mathcal{N}(\Gamma\rightarrow u)+\Gamma-a(M+b)\right]
\end{eqnarray}
\textit{We, then, substitute the expression for Normal distribution given by $\mathcal{N}(\Gamma\rightarrow u)=\frac{1}{\Gamma\sqrt{2\pi}}e^{-\frac{1}{2}\left(\frac{u-\Gamma}{\Gamma}\right)^2}$ to the equation \eqref{nh} to obtain the expression for $\mathcal{H}$. At the $M\rightarrow\infty$ limit, $\displaystyle\sum_{u=0}^{M}\rightarrow\int_{0}^{\infty}$. Then, after integration and rearranging the terms, we have,}
\begin{eqnarray}
\label{nh1}
\mathcal{H}=e^{-a(M-b)+\Gamma}\left[\Gamma+\frac{d_1ln(\Gamma\sqrt{2\pi})+d_2}{\Gamma\sqrt{2\pi}}-a(M+b) \right]
\end{eqnarray}
\textit{where, $d_1=\sqrt{\frac{\pi}{2}}+\sqrt{2}erf\left(\frac{1}{\sqrt{2}}\right)$ and $d_2=\frac{1}{2}erf\left(\frac{1}{\sqrt{2}}\right)-\frac{e^{-\frac{1}{2}}}{\sqrt{\pi}}$ are constants respectively. $erf(x)$ is the error function of $x$. From equations \eqref{I} and \eqref{nh1} and putting the value of $\mathcal{H}_{max}=e^{-a(M+b)+\Gamma}\times ln(M)$, we have the information stored in classical Normal pattern, $\mathcal{I}$ as in the following,}
\begin{eqnarray}
\label{nh2}
\mathcal{I}=e^{-a(M-b)+\Gamma}\left[ln(M)+a(M+b)-\Gamma-\frac{d_1ln(\Gamma\sqrt{2\pi})+d_2}{\Gamma\sqrt{2\pi}} \right]
\end{eqnarray}
\textit{In order to have non-negative information, one has to have the condition $ln(M)+a(M-b)>\Gamma+\frac{d_1ln(\Gamma\sqrt{2\pi})+d_2}{\Gamma\sqrt{2\pi}}$. We further get that at the sufficiently large time, $\displaystyle\mathcal{I}_\infty=\lim_{t\rightarrow\infty}\mathcal{I}=e^{b}\displaystyle\left[ln(M)-b-\frac{d_1ln(b\sqrt{2\pi})+d_2}{b\sqrt{2\pi}} \right]$. This indicates that at large time limit, the overall production rate of cytokine should be controlled such that $ln(M)\gg b$ as the third term is less than one, otherwise, $\mathcal{I}_\infty<0$ which is physically have no meaning.
}

\section{Conclusion and discussion}
{\noindent}Patterns are manifestation of regular self-organised aggregated random components. They have significant roles as they keep diverse information stored in them to take part in collective decision making. Hence, it is important to study these patterns and information associated with them specially in biological systems. \\

{\noindent}The interaction of cytokine and T-cells in our body keeps the regulation of immune system and helps in various biological functions \cite{Wu}. Since the cytokine production by T-cells is quite essential for developing immune system, we considered a simple of cytokines and T cells interaction, where, all possible cytokine production mechanisms are associated. We used stochastic formalism to study the model by constructing Master equation of the model system specifically considering the dynamics of cytokines. The constructed Master equation is solved analytically using generating function technique and the solved probability distribution function, which describes the pattern of distribution of dynamical cytokines in the system, is found to be quite complex. However, under certain limiting conditions, we explored few interesting patterns which carry the information of the system. Near thermodynamics limit, the probability distribution function becomes modified Poisson distribution which indicates that the distribution of the cytokines in the system exhibit classical Poisson pattern. This property indicates that the cytokine events occurring in the system is purely random and the pattern is exhibited mostly by the self-organised aggregation of spatiotemporal random cytokine events scattered over the entire system. In such situation, Poisson pattern behaves as a basin of attraction which attracts cytokine random processes towards it and nonrandom processes are away from it \cite{Frank}.  The uncertainty or surprise of random cytokine events is measured by complexity measurement parameter Shannon entropy which we found to be time dependent. The information stored in the Poisson pattern is then calculated as the difference of maximum entropy and entropy at any instant of time, and found that the competition between overall production rate of cytokines and overall decay rate of it need to be balanced such that information stored in the Poisson pattern should be optimised and non-zero.\\

{\noindent}On the other hand, near the thermodynamics limit, if the mean of the cytokine distribution is taken to be significantly large, then the distribution of the cytokines exhibits classical Normal pattern. In such situation, the phenomenon indicates that the sum of the observations of the small-scale random and independent cytokine events is also a random variable which is a larger-scale random process in the system, no matter how small the distributed random variables are \cite{Lyon}. The distribution of such larger-scale random processes in the system approaches to classical Normal pattern. It also reveals that not only the small-scale random processes, but also larger-scale random processes of the cytokine distribution exhibit Normal pattern. This could be an indication of hierarchical organization of the cytokine random processes which approaches to similar behaviour at various scales. The information contained in the Normal pattern is calculated using Shannon entropy of the distribution.\\

{\noindent}The role of the noise is important and significant specially in microscopic systems. In the cytokine and T-cells interaction model, the strength of the noise is measured using Fano factor which can be calculated from the information of mean and variance of the distribution. The calculated Fano factor shows noise enhanced or super-Poissonian process which indicates that the noise associated with the distribution can significantly drive the system at various possible accessible states. In such situation, the state of the cytokine distribution and dynamics is driven far away from the equilibrium and are generally in active state.\\

{\noindent}The study in this simple model could able to understand few important features and properties of cytokines distribution which is dynamic in principle in cytokine and T-cells interaction. This system is multi-variable system which is needed to incorporate various other interacting molecular species. In such situation, analytical work could have limitation but definitely large scale simulation can be implemented to the system to study dynamics and distribution of the cytokine and related other variables. \\

\vspace{0.5cm}
\noindent {\bf Acknowledgements} \\
{\noindent}RKBS is financially supported DST-Matrics scheme.\\

\noindent {\bf Author Contributions:}\\
{\noindent}RKBS conceptualized the work. MSS, MKS and RKBS did the analytical work. All authors read, analyzed the results, and approved the final manuscript.\\

\noindent {\bf Additional Information} \\
\textbf{Competing interests:} \\ The authors declare no competing interests.

\end{document}